\documentclass[12pt]{iopart}
\usepackage{graphicx}
\usepackage{float}
\usepackage{subfig}
\usepackage{setspace}
\usepackage{color}
\usepackage{multirow}

\begin{document}
\title{Real-time planar flow velocity measurements using an optical flow algorithm implemented on GPU}
\author{N. Gautier and J.-L. Aider}
\address{PMMH, 10, rue Vauquelin 75006 Paris, France}

\begin{abstract}
This paper presents a high speed implementation of an optical flow algorithm which computes planar velocity fields in an experimental flow. Real-time computation of the flow velocity field allows the experimentalist to have instantaneous access to quantitative features of the flow. This can be very useful in many situations: fast evaluation of the performances and characteristics of a new setup, design optimization, easier and faster parametric studies, etc. It can also be a valuable measurement tool for closed-loop  flow control experiments where fast estimation of the state of the flow is needed. The algorithm is implemented on a Graphics Processor Unit (GPU). The accuracy of the computation is shown. Computation speed and scalability of the processing are highlighted along with guidelines for further improvements. The system architecture is flexible, scalable and can be adapted on the fly in order to process higher resolutions or achieve higher precision. The set-up is applied on a Backward-Facing Step (BFS) flow in a hydrodynamic channel. For validation purposes, classical Particle Image Velocimetry (PIV) is used to compare with instantaneous optical flow measurements. The important flow characteristics, like the dynamics of the recirculation bubble computed in real-time, are well recovered. Accuracy of real-time optical flow measurements is comparable to off-line PIV computations. \\\\

\end{abstract}


\section{Introduction}
Optical measurements of 2D-velocity fields in fluid mechanics have been widely used in industrial and academics laboratories for the past thirty years. They allow for the thorough investigation of flow physics through non intrusive means and are an invaluable tool for understanding the dynamics of complex flows. The classical measurement technique is the standard 2D2C PIV (measurements of the 2-Components of the velocity field in a 2D plane) \cite{Adrian2005}. It consists in illuminating the seeded flow with a plane laser sheet (typically generated by a pulsed YaG laser) and acquiring two images of the illuminated particles field at two successive time steps using 15 Hz double-frame cameras or fast cameras for time-resolved (1 kHz) measurements. Usually, a few hundreds pairs of images are acquired. In these standard PIV setups, data are transferred or stored on the computer and post-processed off-line because the computations to obtain a well defined velocity field with a good spatial resolution (typically a $16 \times 16$ cross-correlation window) are time-consuming.

The development of reliable, flexible, accurate and low-cost systems capable of computing flow velocity fields in real-time would be a great step forward for the fluid mechanics community. In addition to saving a lot of time and resources, it would allow academics and industrial researchers to visualize the flow velocity field directly and make adjustments to their experiments on the fly. Accurately targeted measurement campaigns would become feasible even for flows exhibiting high frequency behaviors, like flows downstream a bluff body  \cite{Pastoor2008,Joseph2012},  a cylinder \cite{Roshko1961}, or a wing \cite{Hetrick2004}.\\

Furthermore such  systems would open new perspectives for closed-loop flow control experiments based on visual informations instead of wall-pressure or skin friction measurements. For instance,  \cite{Henning2007} used $4 \times 15$ microphones in parallel rows to measure pressure fluctuations downstream of a step. Using quantitative visual informations would be equivalent to mapping the flow with as many captors as the image size divided by the spatial resolution the 2D velocity field.  Visual servoing in flow control has already been suggested and successfully implemented in numerical simulations \cite{Collewet2011}. One can find experimental demonstrations on improvement of the aerodynamic properties of micro air vehicles \cite{Willert2010}, or control of the flow behind a flap \cite{Roberts2012}. Achieving increased performances would allow for additional means of control, such as vortex tracking or slope-seeking \cite{Henning2005}. 
Several approaches have been suggested to achieve real-time PIV. For instance, a bare bones PIV algorithm has been implemented by \cite{Willert2010} on a single processor,  obtaining engaging performances, while \cite{Roberts2012} implements a basic PIV algorithm on a GPU. However these approaches led to velocity fields from small images at relatively low frame rates (less than 20 fps). Direct cross-correlation PIV, and particle tracking velocimetry (PTV) algorithms have been programmed into Field Programmable Gate Arrays (FPGA) \cite{Jacquet2003,Tadmor2005,Liberzon2010}. However a specific Hardware Description Language (HDL) is required to successfully operate them, which is a strong limitation.
The spectacular increase in computing power of GPUs (the peak GFLOPS performance roughly doubles every year)  allows for an alternative means of achieving real-time processing. Indeed, the processing power of graphics cards has risen at a rate superior to that of Central Processing Units (CPU, doubles every two years). Until recently, it was difficult for the layman to access that power for something other than specific applications. With the introduction of GPU extensions for mainstream computing languages (C/C++, Fortran, Python, Matlab) implementing GPU code in a flexible manner has become accessible to the general public.

In the present experimental study, a dense optical flow algorithm developed by \cite{Champagnat2005} was used. Its characteristics and performances in comparison to PIV algorithms are comprehensively detailed in \cite{Plyer2011} and \cite{Sant2009}. One the advantages of this algorithm is its scalability. Performance of the algorithm increases hand in hand with GPU computing power.
While optical flow algorithms have been used before to compute flow velocity fields they have never, to our knowledge, been implemented in a real time setup. A comprehensive overview of the algorithms used to compute flow velocity fields can be found in \cite{Schnorr2009}.

\begin{figure}[H]
\centering
\includegraphics[width=0.75\textwidth]{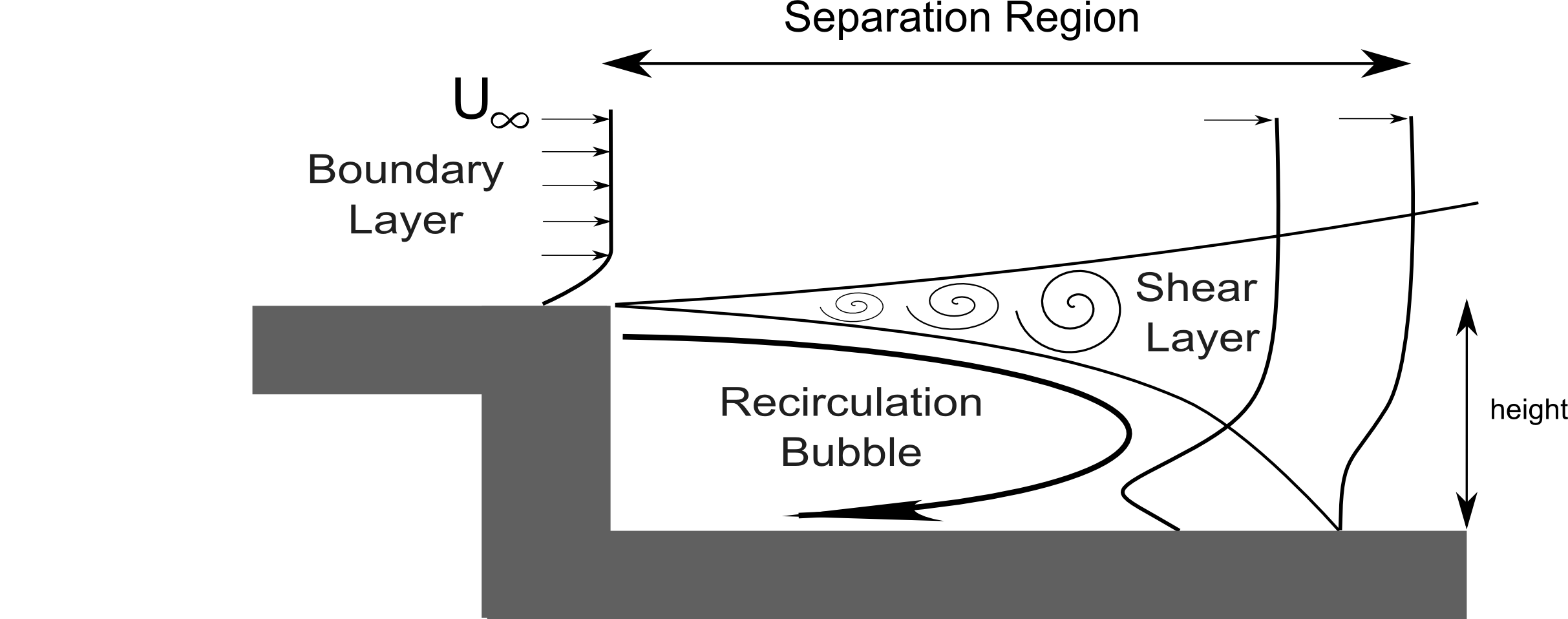}
\caption{Sketch of the backwards facing step flow}
\label{fig:sketchBFS}
\end{figure}

Furthermore a traditional PIV setup can  be cheaply upgraded to a real time PIV setup. In addition to the efficiency of this algorithm, one should also emphasize its robustness: it can be applied on various active or passive scalar fields, like thermal or dye scalar field. It is then much more flexible and can be applied in much more various experimental situations than standard PIV measurements restricted to particle tracking. 
To demonstrate the efficiency and quality of real-time velocity computations,  it has been tested on a backward-facing step flow. Boundary layer separation and reattachment occur in many natural and industrial systems, such as diffusors, combustors or external aerodynamics of ground or air vehicles. The backward-facing step is the simplest geometry to study a separated flow. Though the geometry is simple, the complexity of separated flows is recovered as shown in figure \ref{fig:sketchBFS}. In this case, the separation is imposed by a sharp edge, allowing for the separation-reattachment process to be examined by itself. A dominant, global feature of the flow is the creation of a large recirculation bubble downstream the step edge, as shown in figure \ref{fig:sketchBFS}. This flow has been extensively studied through experimental and numerical investigations, see \cite{Armaly1983,Chun1996,Le1997,JLA2004,JLA2007}. As the objective of the present paper is exclusively the experimental demonstration of high-speed, efficient and reliable real-time velocity measurements, the BFS flow characteristics will not be discussed thoroughly but solely used as valuable benchmark for this experimental technique.

\section{Experimental Setup}
\subsection{Water tunnel}
Experiments were carried out in a hydrodynamic channel in which the flow is driven by gravity.
 The walls are made of Altuglas for easy optical access from any direction. The flow is stabilized by divergent and convergent sections separated by honeycombs. The test section is $80$~$cm$ long with a rectangular cross section $15$~$cm$ wide and $10$~$cm$ high.\\
 The mean free stream velocity $U_{\infty}$ ranges between $1.38$ to $22$~$cm.s^{-1}$.  The quality of the main stream can be quantified in terms of flow uniformity and turbulence intensity.  The standard deviation $\sigma$ is computed for the highest free stream velocity featured in our experimental set-up. We obtain $\sigma = 0.059$~$cm.s^{-1}$ which corresponds to turbulence levels $\frac{\sigma}{U_{\infty}}=0.0023$.

\subsection{Optical flow measurement set-up}
The flow is seeded with 20~$\mu m$ polyamid seeding particles.  The vertical middle plane of the test section is illuminated from above (figure \ref{fig:SketchAq}) by a laser sheet created by a 2W continuous CW laser operating at a wavelength $\lambda = 532$~nm. 

 The pictures of the illuminated particles are recorded using a relatively low cost (compared to double-fame or high-speed cameras traditionally used for PIV), Basler acA 2000-340km 8bit CMOS camera,with a maximum bandwith of 680 Mb/s. Its resolution is 2048 $\times$ 1088 pixels. The maximum frame rate for full-frame acquisition is $F_{acq} = 340$~$Hz$. The camera is controlled by a camera-link NI PCIe 1433 frame grabber allowing for real-time acquisition and processing. It should be noted that CPU performance is irrelevant with regards to the performance of the optical flow algorithm which runs entirely on the GPU.  In our set-up, a NVIDIA Gforce 580 GTX GPU card, with 520 processing cores clocked at 800 Mhz, has been used. A complete description of this GPU's architecture can be found in \cite{whitepaper500}. The data flow for the acquisition apparatus is detailed in figure \ref{fig:SketchAq}. 
 The images can either be written to a solid state drive or computed in real-time on the GPU. Usually no data is written during visualization of velocity fields to improve performance and frequency rate of the computation. The optical flow algorithm and camera acquisition software are integrated into a single interface using LabView. 
It is important to emphasize the only requirement to upgrade a classic PIV setup featuring a camera streaming images to an acquisition computer, to a setup capable of real-time flow velocity computations is adding a graphics card to the acquisition computer. Therefore this can be done cheaply and with minimal effort.

\subsection{Backward-facing step geometry}
\begin{figure}[H]
\centering
\includegraphics[width=1.0\textwidth]{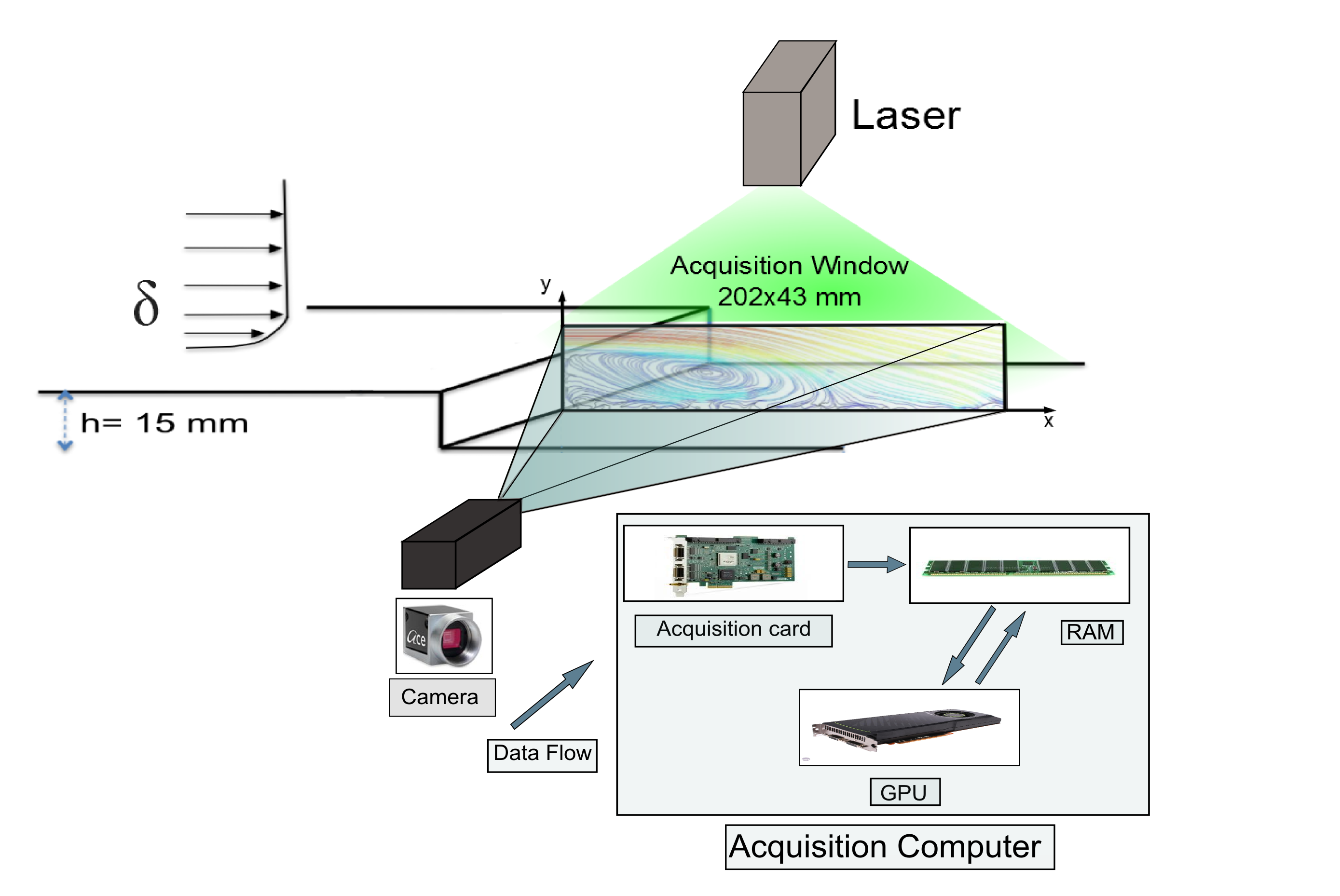}
\caption{Sketch of the backward-facing step and data flow for the acquisition apparatus}
\label{fig:SketchAq}	
\end{figure}

The backward-facing step geometry is shown in figure \ref{fig:SketchAq}. A specific leading-edge profile is used to smoothly start the boundary layer which then grows downstream along the flat plate, before reaching the edge of the BFS. The boundary layer has a shape factor $H \approx 2$. Step height $h$ is 15mm allowing for a range of Reynolds numbers  $0 <Re_h = \frac{U_{\infty}h}{\nu} < 3000$, $\nu$ being the kinematic viscosity. 

\subsection{Optical flow algorithm}

Optical flow is related to the domain of image motion or optical flow estimation in computer vision. This particular algorithm called FOLKI was written in C++/CUDA. It was developed, implemented and rigorously validated by \cite{Plyer2011} at ONERA. To achieve optimal performances further improvements were made by improving memory transfers and enhancing kernel concurrency. A guide on CUDA programming is available at \cite{cudaProg}.  This algorithm was used by \cite{Leclaire2012}, \cite{Bur2012}, and \cite{Blanquart2012}. It is a local iterative gradient-based cross-correlation optimization algorithm which yields dense velocity fields, i.e. one vector per pixel.  It belongs to the Lucas-Kanade family of optical flow algorithms \cite{Lucas1984}. It should be noted the dense nature of the output is intrinsically tied to the nature of the algorithm. The spatial resolution however is tied to the window size, like any other window based PIV technique. However the dense output is advantageous since it allows the sampling of the vector field very close to obstacles, yielding good results near walls, as shown in \cite{Plyer2011}. Computing dense fields allows for a highly parallel algorithm which can take full advantage of the GPU architecture. The interrogation window radius $r=10$ pixels  was chosen, following the guidelines given by \cite{Plyer2011}. It should be noted similar performances would be achieved using other programming languages, such as OpenCL in conjunction any GPU's , though the algorithm would need to be tweaked for the specific  GPU architecture. Because the algorithm is based on light intensity displacement between two images, it should be noted that unlike PIV algorithms this algorithm does not require a seeded flow to compute actionable data. For instance, the same setup could be used to measure velocity fields of flows presenting passive scalar (like fluorescent dyes) concentration gradients or even variations in refractive index due to thermal or density variations.

\begin{figure}
\centering
\includegraphics[width=0.40\textwidth]{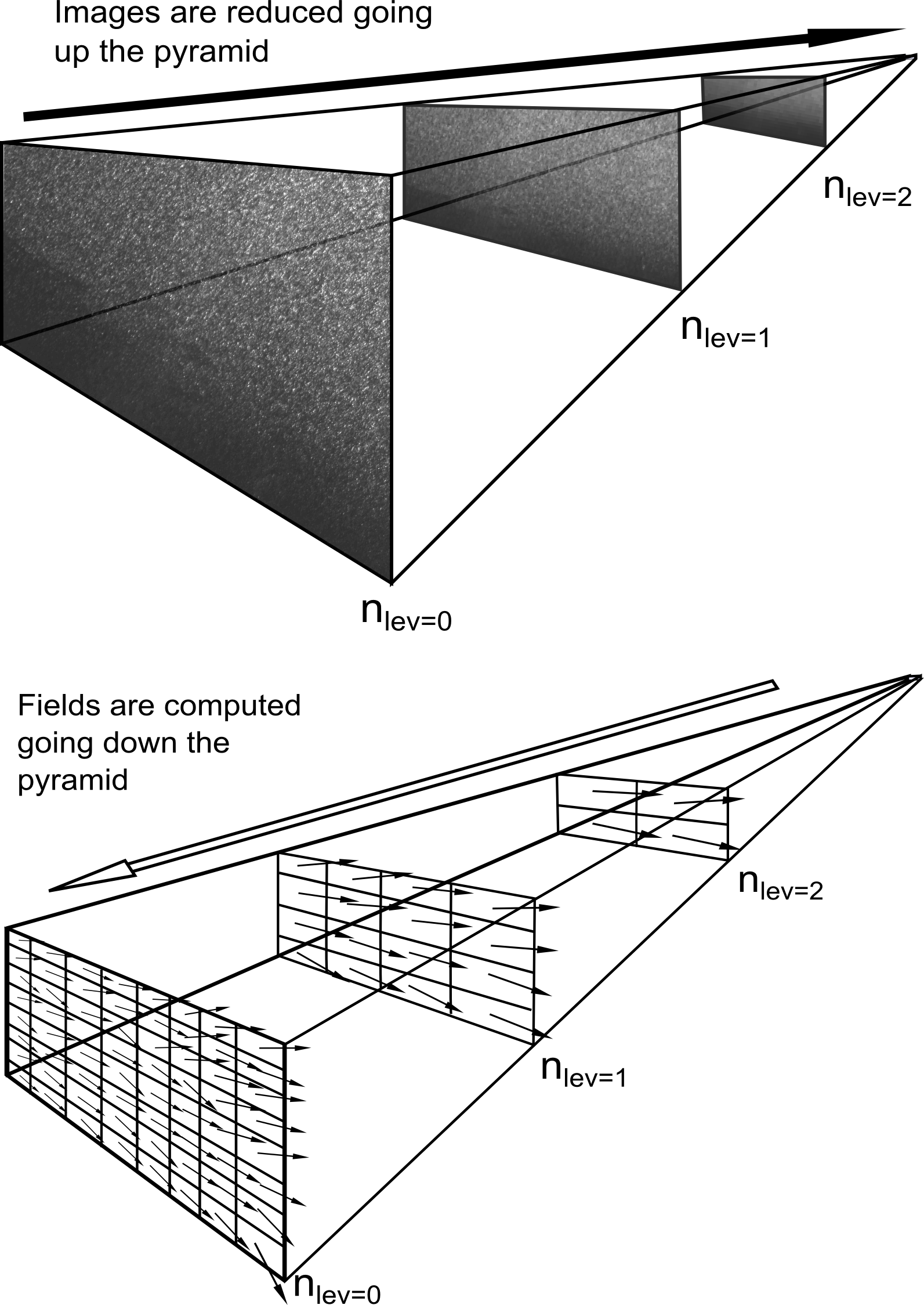}
\caption{Sketch of the computation pyramid.}
\label{fig:pyramid}
\end{figure}

The principle of the featured optical flow algorithm is as follows. The original images are reduced in size by a factor of 4 iteratively until intensity displacement in the reduced image is close to 0. This gives a pyramid of images, described in figure \ref{fig:pyramid}. Displacement is computed in the top image with an initial guess of zero displacement using an iterative Gauss-Newton scheme to minimize a sum of squared difference criterion. This displacement is then used as an initial estimate for the same scheme in the next pair of images in the pyramid. And so on until the base of the pyramid, corresponding to the initial images is reached, thus giving the final displacement. 

The optical setup is tuned for the displacement of the particles to be small enough for the optical flow algorithm to converge. Thus there are two inputs to the algorithm, besides image size, that have a major impact on performance: the number of levels in the pyramid $n_{lev}$, and the number of iterations per level $n_{iter}$ required to achieve convergence of the velocity field. Computing speed is a function of these two integers. If $(\delta x)_{max}$ is the maximum displacement, as a general rule $n_{lev}$ must verify equation \ref{eq:nlevels} from \cite{Plyer2011}:

\begin{equation}
(\delta x)_{lmax} / 2^{n_{lev}-1} < 3 \hspace{1mm}  pixels
\label{eq:nlevels}
\end{equation}

One can see that choosing the time step between images defines the value of $n_{lev}$. One must then choose $n_{iter}$. A low value will give higher performances with slightly lower result quality. When working in real-time a low value (1 or 2) of $n_{iter}$ is recommended. However for off-line computations the value should be raised to ensure full convergence. Performances should still be greater than with commercial PIV software.

While a number of pre and post-processing options are usually used to enhance the computed velocity fields, these operations have a computational cost. Therefore a balance must be found between obtaining actionable data and processing speed. \\

Raw images are pre-processed using a standard local equalization algorithm. This step is implemented on the GPU for increased performance. We have found this step to be mandatory for experimental images processing. Without, the computation does not yield usable data. 

Original image intensity is  normalized  following equation \ref{eq:ZNSSD}:

\begin{equation}
\forall (x,y)\in I, \tilde{I}(x,y)=\frac{I(x,y)-\bar{I}(x,y)}{\sqrt{\bar{I}^2(x,y)-\overline{I^2}(x,y)}}
\label{eq:ZNSSD}
\end{equation}
where $\bar{I}$ is a \textit{local} mean for a given radius. We choose a radius of 5 pixels. This zero normalized sum of square differences (ZNSSD) is common for PIV pre-filtering. The aim is to eliminate the influence of  illumination inhomogeneities. For each pixel the mean intensity value is subtracted (zero mean) and divided by the local intensity standard deviation (normalized). 
\\
The nature of the algorithm is such that computed velocity fields are naturally smooth. Thus there is no post processing required to cull spurious vectors.

For our setup, only a fraction of the camera resolution is used (the region of interest (ROI) $=$ 1792 $\times$ 384). It is enough to capture the whole recirculation bubble downstream the BFS, while ensuring good computing performances. Reducing the time step $\delta t$ between two pictures acquisitions allows lower values of $n_{lev}$ and higher performances. $n_{lev}$ can be lowered to $0$ with a small enough displacement. Decreasing $n_{lev}$ shifts the burden of performance to the camera. $n_{iter}$ should be raised until the computed velocity field does not vary, with $n_{iter} \geq 1$.  $n_{iter} \geq 10$ is seldom needed. With current hardware it is difficult to achieve satisfactory performances with a high number of iterations ($n_{iter} > 4$).  $n_{iter}$ can be brought down as low as 1 and still yield actionable quantitative information on the flow, with a significant improvement in computing times. 

\begin{figure}
\centering
\includegraphics[width=0.75\textwidth]{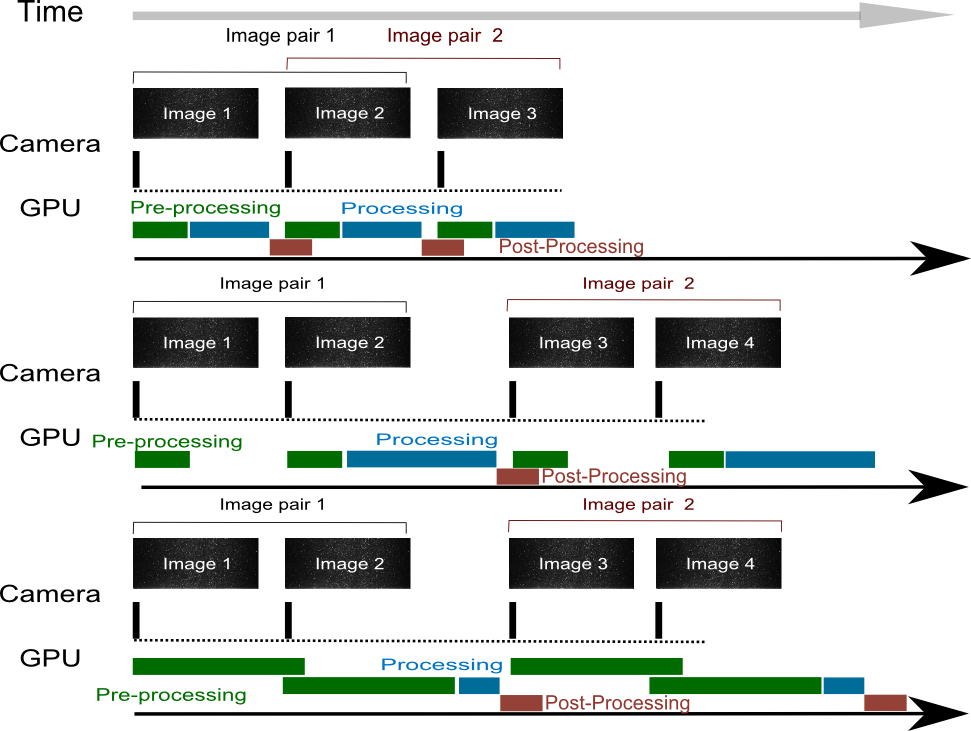}
\caption{Different computing schemes}
\label{fig:schemes}
\end{figure}

Concerning latency, depending on GPU performances and camera acquisition frequency different computing schemes are implemented for optimal performances as show in figure \ref{fig:schemes}.

The first scheme is used when computation is fast enough to keep up with the camera. This is the fastest scheme by far since each field computation requires only one image to be processed. The second is used when the first cannot and pre-processing time is lower than camera exposure. Finally when preprocessing takes too long, preprocessing on the second image starts while preprocessing on the first image is finishing. Latency varies depending on the scheme but is upper bounded by exposure time plus the time required to preprocess one image and compute the corresponding field. Post-processing can also be \textit{hidden} during copy from the camera to the GPU, a period during which the GPU is iddle. Post-processing here refers to the computation of integral quantities from the data.

\subsection{PIV computations}
To validate the optical flow measurements standard PIV algorithms were used off-line. The Davis software from LaVision was used, using a PIV multi-pass cross-correlation algorithm with a final 16$\times$16 pixel interrogation window with 50 \% overlap, thus leading to PIV fields with a 8 $\times$ 8 pixel grid resolution.

\section{Results}

\subsection{Real-time computation of instantaneous 2D velocity fields}

\begin{figure}
\centering
\subfloat[]{\includegraphics[width=0.55\textwidth]{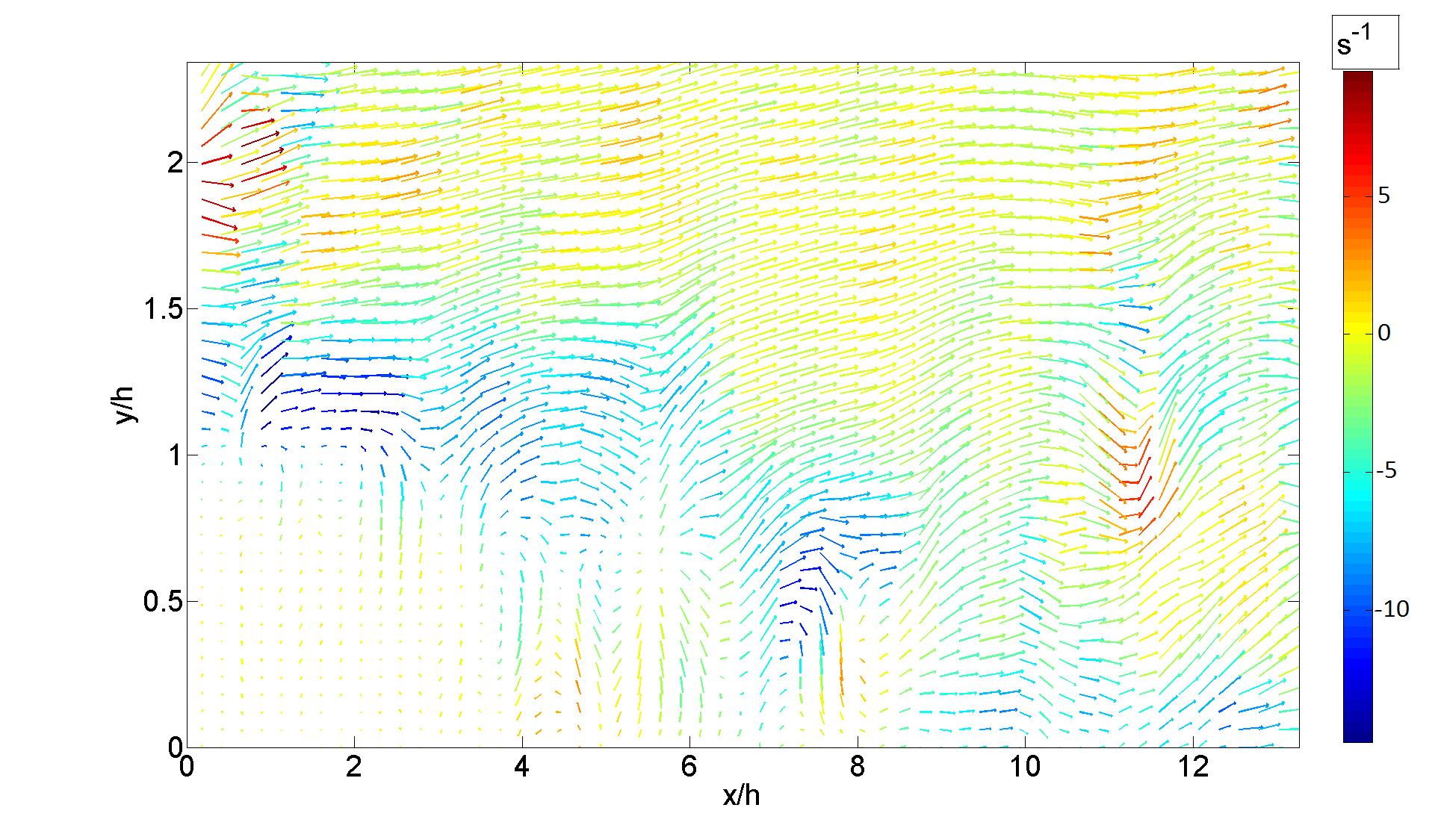}\label{fig:vectorPIV}} \subfloat[]{\includegraphics[width=0.55\textwidth]{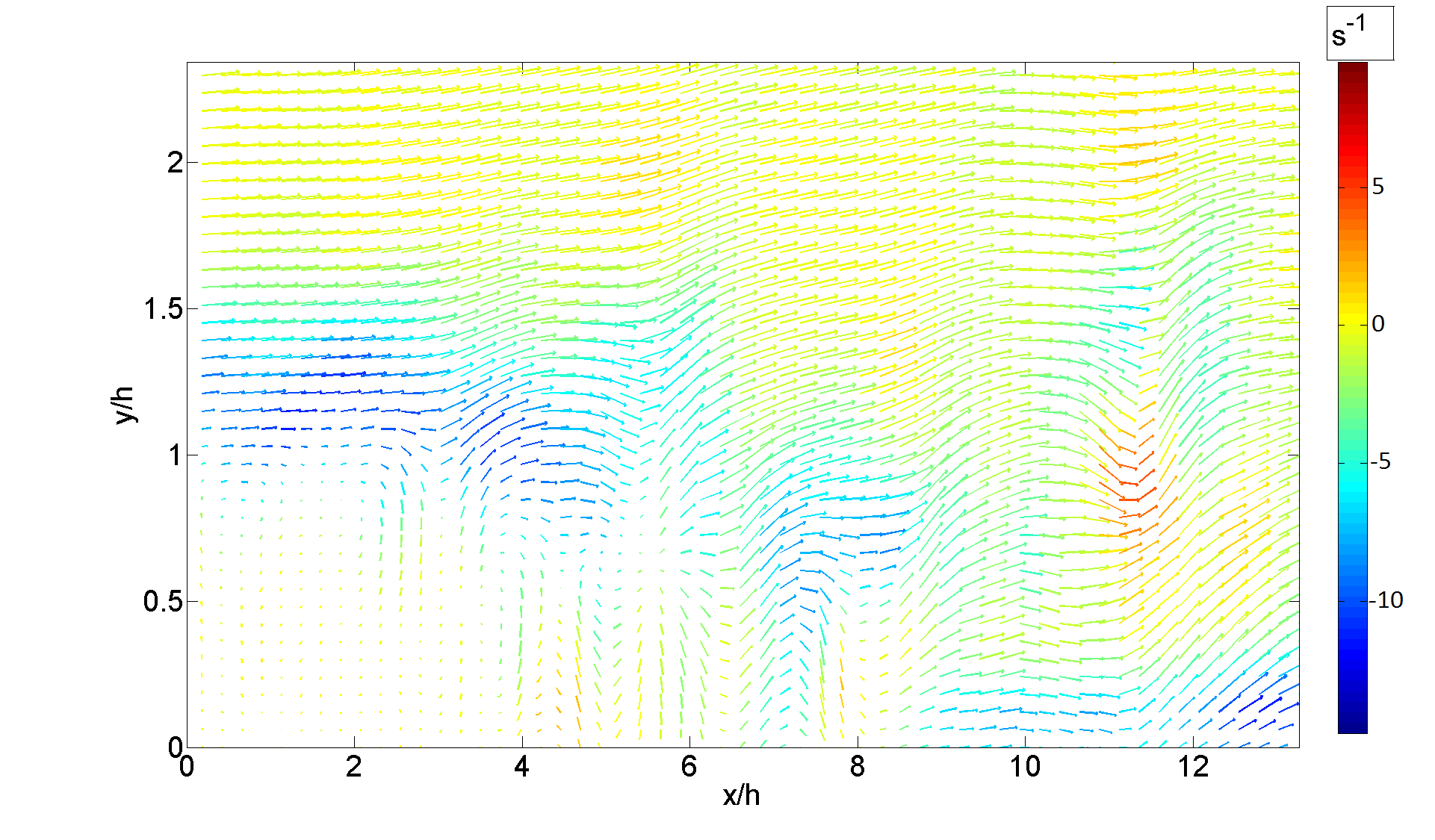}\label{fig:vectorOF}}
\caption{Vector plots of the instantaneous velocity field downstream the BFS for $Re_h=2500$, colored by vorticity, and computed by PIV (a) and by optical flow (b).}
\end{figure}

Figure~\ref{fig:vectorPIV} shows an instantaneous vector plot colored by vorticity for $Re_h=2500$ obtained with a PIV algorithm. Figure~\ref{fig:vectorOF} shows the same velocity fields but computed with our optical flow algorithm. One can see that spanwise Kelvin-Helmholtz vortices are well captured. Differences between the two velocity fields are mainly due to poorer results of the PIV algorithm  near the edges of the acquisition window and near the walls.

\subsection{Comparison of the real-time optical flow measurements with off-line PIV computations}
The accuracy of the algorithm has been demonstrated \emph{off-line} for numerical and experimental  data by \cite{Plyer2011}. In this section we will focus on the computation of an integral scalar value derived in real-time from the instantaneous velocity fields. The objective is twofold: illustrate the real-time computation of a global quantity extracted from instantaneous velocity fields in real-time and evaluate the accuracy of the optical flow computations compared to standard PIV computations. 

There are a number of pertinent integral values which can be used to characterize the separated flow. In this experiments we choose to compute the surface of the recirculation bubble in the instantaneous 2D velocity field. It is straightforward and quick to compute while remaining a good way of evaluating the state of the flow. The recirculation bubble surface is defined as the percentage of velocities below a given threshold $T_v$ relative to the total number of available velocity vectors in the whole velocity field on image $I$. The threshold is chosen as the median between the lowest and highest longitudinal velocity for the mean velocity field. The recirculation bubble surface $S_{bubble}$ is defined by equation \ref{eq:surf}:

\begin{equation}
S_{bubble}=\frac{\sum_{I}{v_x(x,y)\le T_v}}{\sum_{I}{v_x(x,y)}}
\label{eq:surf}
\end{equation}
with $T_v=\frac{1}{2}[\min \bar{v_x}(x,y)+\max \bar{v_x}(x,y)]$.

\begin{figure}
\centering
\includegraphics[width=0.5\textwidth]{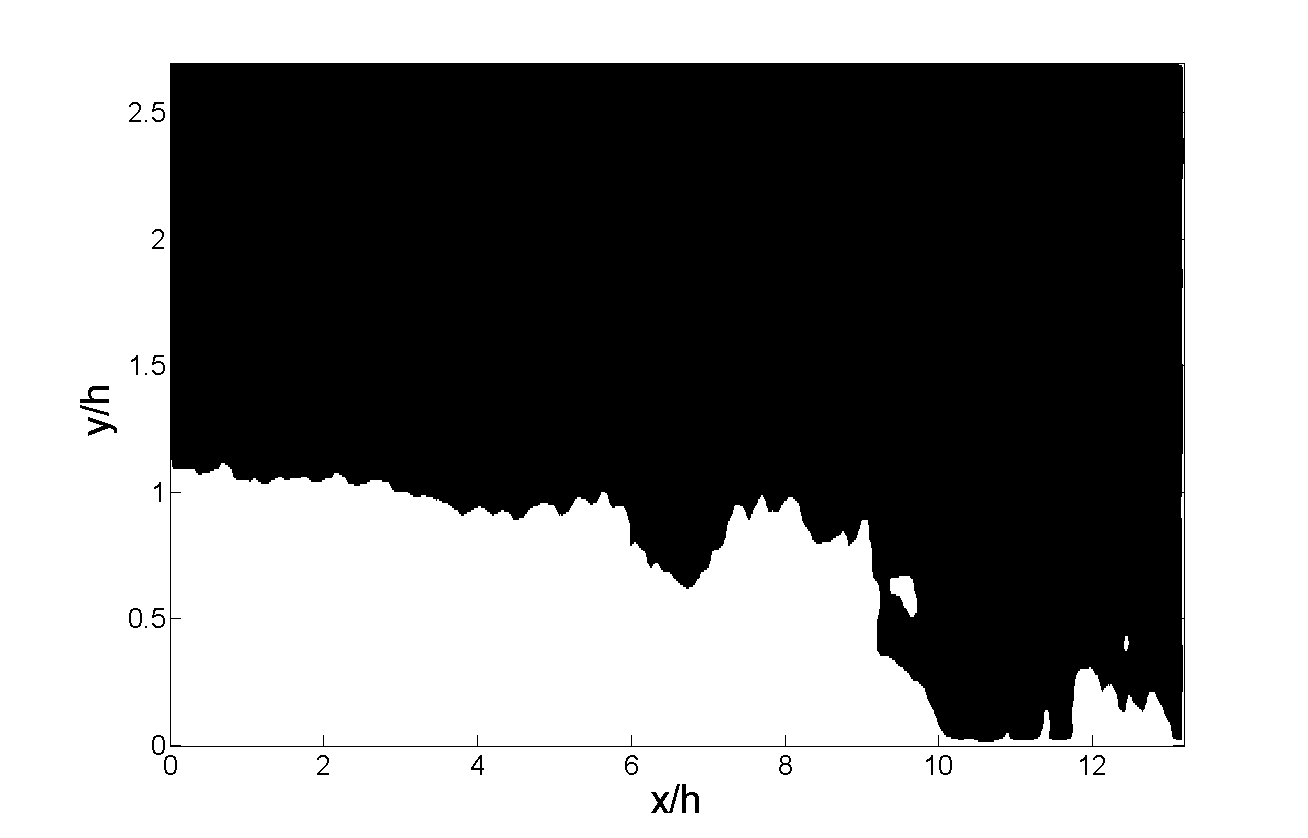}
\caption{Visualization of the recirculation bubble area (white region) defined by a velocity threshold applied on the instantaneous velocity field. In this case, $S_{bubble}$=21\% is computed in real-time.}
\label{fig:vxOFBW}
\end{figure}

Figure \ref{fig:vxOFBW} shows the values below the threshold for an instantaneous velocity field. One can see that the white area gives a good estimate of the instantaneous recirculation bubble. $S_{bubble}$ is the area of the white region given as a percentage of the total velocity field area. Such computations are carried out for both optical flow and PIV velocity fields. The recirculation bubble surface can be correlated to the reattachment length $L_R$ usually used to characterize the BFS flow \cite{Armaly1983,JLA2007}. Such an integral value could, for example, be used as an input in a feedback loop.

\begin{figure}[H]
\centering
\includegraphics[width=0.55\textwidth]{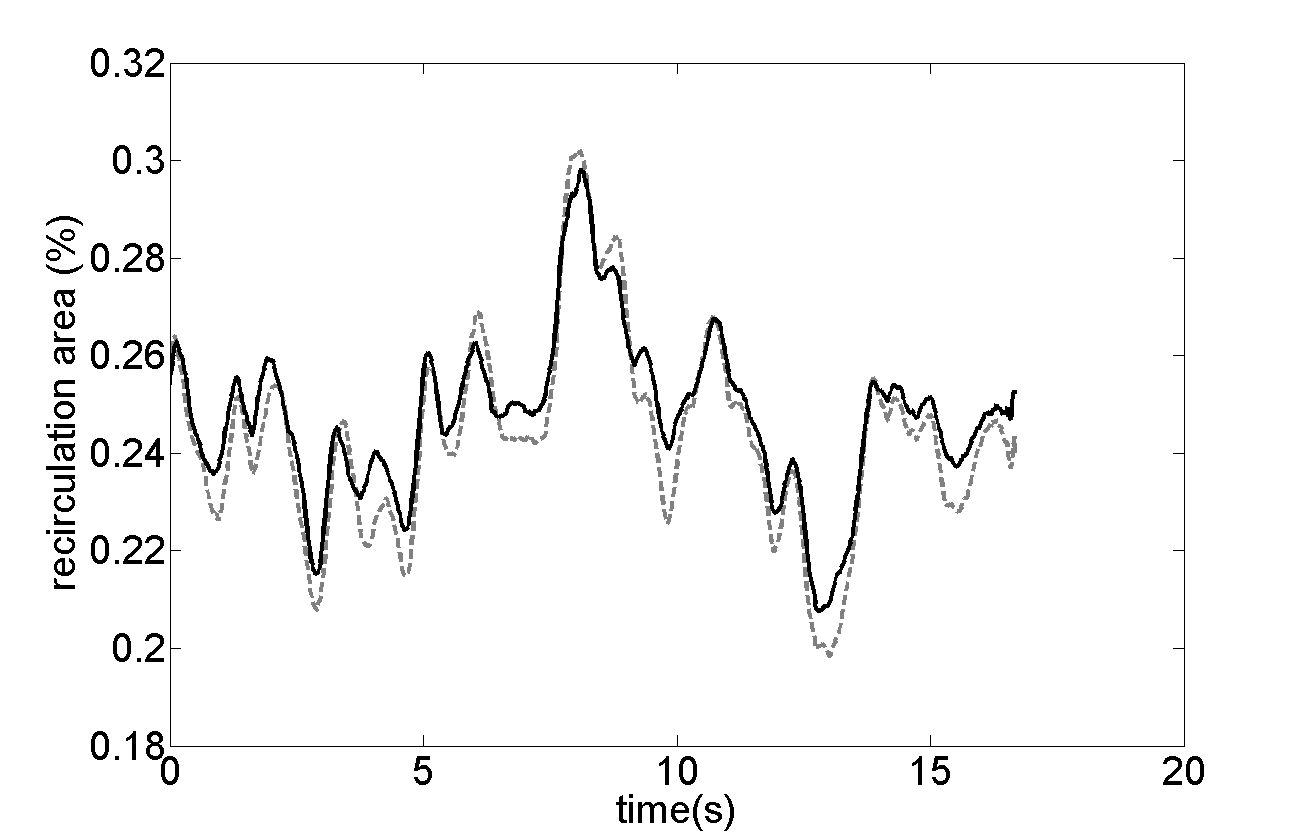}
\caption{Comparison of time series of recirculation bubble surface $S_{bubble}$ computed using PIV (black full line) and optical flow (grey dotted line) for a time-resolved velocity measurement.}
\label{fig:volOFvsPIV}
\end{figure}

Figure~\ref{fig:volOFvsPIV} shows time series of instantaneous $S_{bubble}$ computed with optical flow and with PIV for a time-resolved series computed off-line. Data featured in figure~\ref{fig:volOFvsPIV} are for $Re_h=2500$ and with an acquisition rate of 60 image pairs per second. One can see that there is a good agreement between the two time-series. Differences can be explained by the fact some images do not contain enough particles in the recirculation region. The optical flow and PIV algorithms converge in slightly different ways. Because of the dense nature of the optical flow output, more information is available near the walls \cite{Plyer2011}. The computed recirculation bubble is more clearly defined and is subject to greater variations as shown in figure~\ref{fig:volOFvsPIV}. 

\begin{figure}[H]
\centering
\subfloat[Comparison of  mean $S_{bubble}$ computed with PIV ($\times$ and full line) and optical flow ($\circ$ and dotted line) algorithms.]{\includegraphics[width=0.49\textwidth]{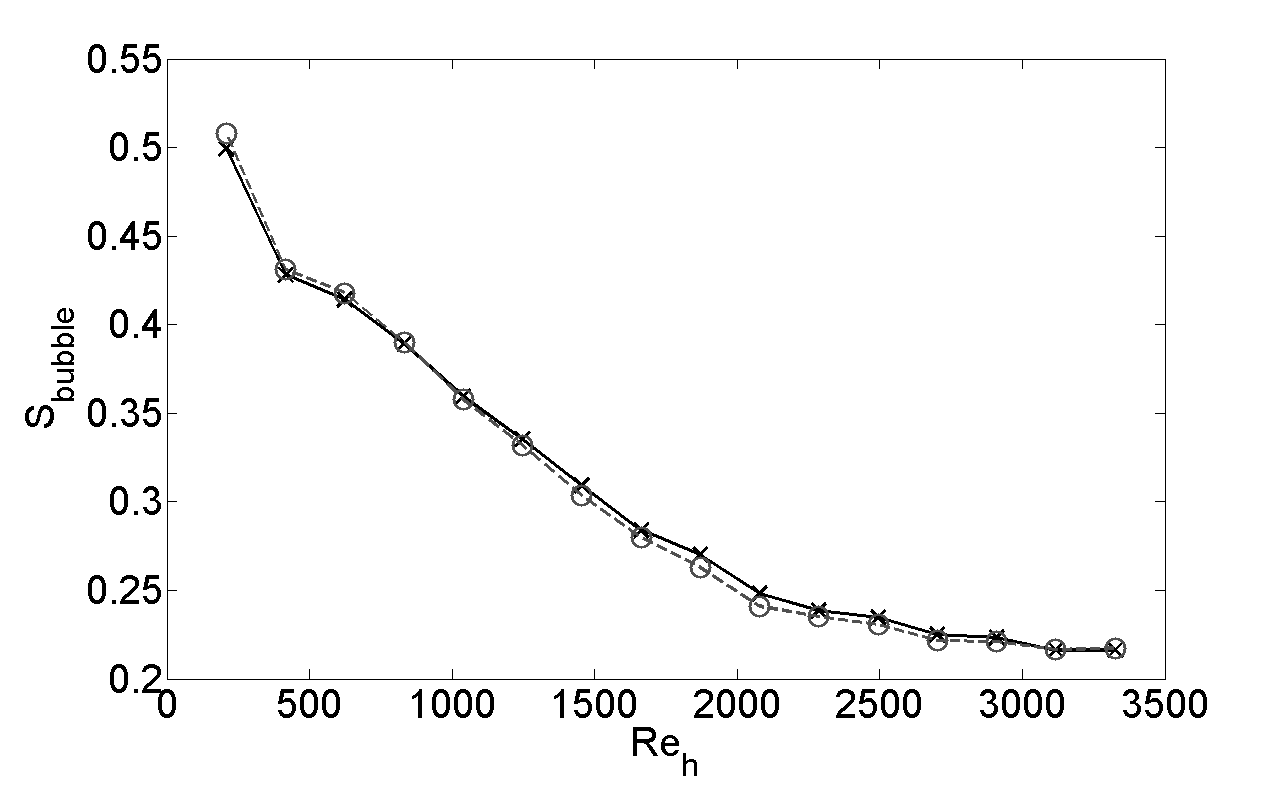}\label{fig:volumePIVvsOF}}
\hspace{1mm}
\subfloat[Relative error on recirculation bubble surface between PIV and optical flow]{\includegraphics[width=0.49\textwidth]{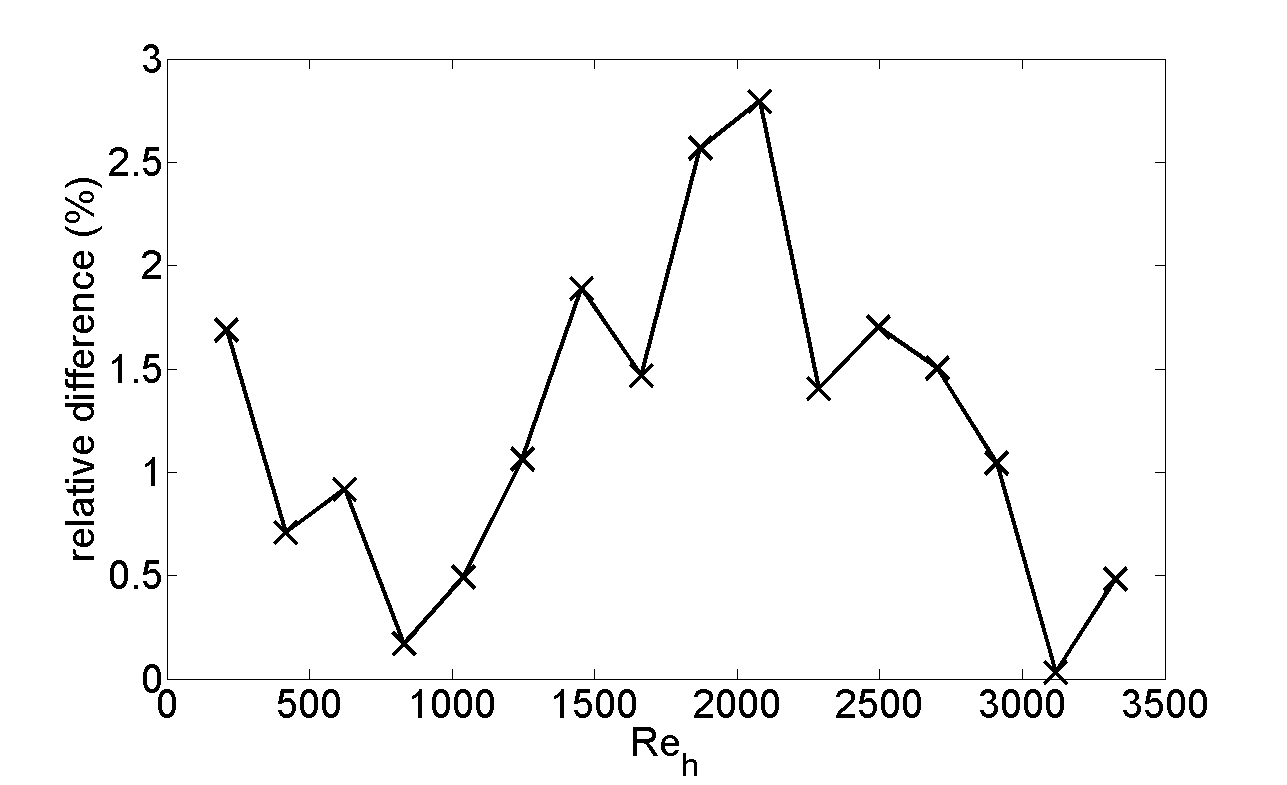}\label{fig:errorPIVvsOF}}
\caption{}
\end{figure}

Figure~\ref{fig:volumePIVvsOF} shows a comparison of the surface of the mean recirculation bubble as a function of Reynolds number computed by PIV or optical flow. Figure~\ref{fig:errorPIVvsOF} shows the relative difference between results obtained with both algorithms. Agreement is good with a relative difference always lower than 3\%. It shows that the optical flow computations is robust over a wide range of Reynolds numbers corresponding to complex  instantaneous flows. 

\subsection{Optimizing the computation frequency}
\begin{table}[H]
\footnotesize{
\begin{tabular}{cc|c|c|c|c|c|c|c|}
\cline{3-9}
&  &\multicolumn{7}{ c| }{$n_{Iter}$} \\ \cline{3-9}
&  & 1 & 2 & 3 & 4 & 5 & 6 & 7 \\ \cline{1-9}
\multicolumn{1}{ |c|   }{\multirow{4}{*}{$n_{lev}$} } &
\multicolumn{1}{ |c| }{\multirow{2}{*}{3}} & 4.80\% & 2.12\%& 1.45\%& 1.14\%& 1.00\% & 0.99 \% & 0.2 \%    \\
\multicolumn{1}{ |c  }{}                        &
\multicolumn{1}{ |c| }{} & 224.0 fps & 148.0 fps& 112.0 fps & 89.3 fps& 73.5 fps & 63.0 fps & 55.7  fps    \\ \cline{2-9}
\multicolumn{1}{ |c  }{}                        &
\multicolumn{1}{ |c| }{\multirow{2}{*}{4}}& 3.65\% &  2.12\% & 1.45\% & 1.14\%& 1.00\% & 0.93\% &  0\% \\
\multicolumn{1}{ |c  }{}                        &
\multicolumn{1}{ |c| }{}& 220.0 fps & 147.0 fps & 110.1 fps &87.2 fps & 72.4 fps & 62.0 fps &   55.6 fps     \\ \cline{1-9}
\end{tabular}
} 
\caption{Error and fields per second (fps) as a function of computing parameters.}
\label{tab:fpsvserror}
\end{table}

This setup allows for accurate computations of the 2D velocity fields in real-time.  If the aim of the user is real time visualizations and/or quick computations of flow properties for a feedback loop, constraints on the algorithm can be relaxed. Indeed, some compromise can be found if increased computations speed is called for. If the aim is the computation of accurate velocity fields, constraints should be increased to achieve maximum accuracy, while still retaining fast computing times in real time. 

Table \ref{tab:fpsvserror} shows how one can drastically improve computation speed, by lowering the number of levels and iterations while still retaining a meaningful integral information from the flow.

A video is linked to this article showing flow images and real-time computed fields as well as $S_{bubble}$ history.
Higher fields per second can be achieved by shortening the time step between two images in order to bring down maximum displacement and thus allowing $n_{lev} \rightarrow 0 $.  This translates into a maximum of 350 fps at this resolution. Higher speeds can be achieved by reducing the field of view, for example by focusing on an area where fluctuations are most present or using masks to avoid computations over obstacles or side walls. 
\\
The aforementioned performances are achieved with a given hardware. The scalability of the algorithm ensures greater performance with better GPUs. Moreover since the computation of a velocity field is independent of all other computations, adding additional graphic cards to the setup would allow for a proportional increase in computation speed. The only limit to the achievable frame rate are the acquisition rate of the camera image quality. For very high acquisition frequencies, a more powerful CW laser, a pulsed YaG laser and/or a more sensitive camera are required.

\section{Conclusion and perspectives}
We have shown how a simple and relatively low cost setup can be configured to achieve high speed real-time computations of a flow velocity field and computation of relevant values from this velocity field. The key feature of the setup is the use of an optical flow algorithm which takes advantage of the massively parallel processing capabilities of GPUs. This work is of use to experimentalists who wish to observe flow properties in real-time as well as those who wish to use high frequency flow data to implement a feedback-loop in flow control experiments. We have demonstrated the accuracy of the method by comparing our results with results obtained by the more widely used PIV approach, computed off-line. Finally, ways of improving computing speed and reaching higher frame rates are discussed. 
\section{Acknowledgments}
The authors gratefully acknowledge the support of the DGA, as well as Aurelien Plyer for all his helpful advice.

\bibliographystyle{unsrt}	
\bibliography{Bibliography}

\end{document}